\documentclass[twocolumn,showpacs,amsmath,amssymb]{revtex4}
\usepackage{graphicx}% Include figure files
\usepackage{dcolumn}% Align table columns on decimal point
\usepackage{bm}% bold math
\usepackage{psfig}% bold math
\newcommand{\bip}{B^+}

\newcommand{\biz}{B^0}

\newcommand{\dem}{D^-}

\newcommand{\dzb}{\overline{D}^0}
\newcommand{\dszb}{\overline{D}^{*0}}
\newcommand{\pip}{\pi^+}
\newcommand{\pim}{\pi^-}
\newcommand{\piz}{\pi^0}
\newcommand{\rop}{\rho^+}
\newcommand{\rom}{\rho^-}
\newcommand{\roz}{\rho^0}

\newcommand{\kap}{K^+}
\newcommand{\kam}{K^-}

\newcommand{\kaz}{K^0}
\newcommand{\ko}{K_0}
\newcommand{\ks}{K_S^0}
\newcommand{\kzb}{\overline{K}^0}

\newcommand{\etap}{\eta^{\prime}}
\newcommand{\etac}{\eta_c}

\newcommand{\psp}{\psi^{\prime}}

\newcommand{\jpsi}{J/\psi}

\newcommand{\nnb}{n\overline{n}}

\newcommand{\ppb}{p\overline{p}}

\newcommand{\LLb}{\Lambda \overline{\Lambda}}

\newcommand{\bppb}{(p\overline{p})}
\newcommand{\bllb}{(n\overline{\Lambda})}
\newcommand{\bnpb}{(n\overline{p})}
\newcommand{\bnbp}{(\overline{n} p)}
\newcommand{\bnlb}{(n\overline{\Lambda})}
\newcommand{\bnbl}{(\overline{n} \Lambda)}
\newcommand{\bplb}{(p\overline{\Lambda})}
\newcommand{\bpbl}{(\overline{p} \Lambda)}
\newcommand{\ra}{\rightarrow}

\newcommand{\beq}{\begin{equation}}
\newcommand{\eeq}{\end{equation}}
\newcommand{\beqn}{\begin{eqnarray}}
\newcommand{\eeqn}{\end{eqnarray}}
\newcommand{\beqns}{\begin{eqnarray*}}
\newcommand{\eeqns}{\end{eqnarray*}}
\newcommand{\bfg}{\begin{figure}}
\newcommand{\efg}{\end{figure}}
\newcommand{\bitm}{\begin{itemize}}
\newcommand{\eitm}{\end{itemize}}
\newcommand{\bnum}{\begin{enumerate}}
\newcommand{\enum}{\end{enumerate}}
\newcommand{\btbl}{\begin{table}}
\newcommand{\etbl}{\end{table}}
\newcommand{\btbu}{\begin{tabular}}
\newcommand{\etbu}{\end{tabular}}
\def\Journal#1#2#3#4{{#1} {\bf #2}, #3 (#4)}
\def\CTP{Commun. Theor. Phys.}

\def\NP{Nucl. Phys.}

\def\PLB{Phys. Lett. B}
\def\PRL{Phys. Rev. Lett.}
\def\PR{Phys. Rev.}

\def\PRD{Phys. Rev. D}

\def\PTP{Prog. Theor. Phys.}

\begin{document}
\preprint{Draft-PRD}

\title{\boldmath Baryon Antibaryon Nonets}
\author{C.~Z.~Yuan}
 \email{yuancz@mail.ihep.ac.cn}
\author{X.~H.~Mo}
\author{P.~Wang}
\affiliation{Institute of High Energy Physics, CAS, Beijing
100039, China}

\date{\today}

\begin{abstract}
The baryon-antibaryon $SU(3)$ nonets are proposed as a scheme to
classify the increased number of experimentally observed
enhancements near the baryon antibaryon mass threshold. The scheme is
similar to the Fermi-Yang-Sakata model, which was put forth
about fifty years ago in explaining the mesons observed at that time.
According to the present scheme, many new baryon-antibaryon bound
states are predicted, and their possible productions in quarkonium decays
and $B$ decays are suggested for experimental search.
\end{abstract}


\maketitle

\section{Introduction}

Low mass baryon-antibayron enhancements have recently been
observed in charmonium and $B$ decays. 
In charmonium decays,
$\ppb$ and $p\overline{\Lambda}$ enhancements are observed in
$\jpsi \ra \gamma \ppb$~\cite{besgppb}, 
$\psp \ra \piz \ppb,~\eta \ppb$~\cite{bespaeppb}, and $\jpsi \ra p
\overline{\Lambda} K^- +c.c.$~\cite{bespkl}, as well as in $\psp
\ra p \overline{\Lambda} K^- +c.c.$ decays~\cite{bespkl} by BES
collaboration. In $B$ decays, many baryon-antibayron-pair-contained
final states have been measured by CLEO, Belle and BABAR collaborations,
such as $\biz \ra D^{*-} p \overline{n}$~\cite{cleobdcy}, 
$B^{\pm} \ra \ppb K^{\pm}$~\cite{belbppk}, 
$\overline{\biz} \ra D^{*0} \ppb,~D^{0} \ppb$~\cite{belbdpp}, 
$\biz \ra p \overline{\Lambda} \pi^-$~\cite{belbplpi}, 
$\bip \ra \ppb \pi^+,~\ppb K^{*+}$, $\biz \ra \ppb \kaz$~\cite{belbpppk}, 
$\bip \ra \LLb \kap$~\cite{belbllk}, 
$\biz \ra \overline{D}^{*0} \ppb,~\overline{D}^{0} \ppb$~\cite{bababdcy}, 
and so on, with observed enhancements in $\ppb$, $p\overline{\Lambda}$ and $\LLb$ mass spectra. Except for the enhancement in $\jpsi \ra
\gamma \ppb$, which is claimed to be very narrow and below the
$\ppb$ mass threshold, all other states are slightly above the
baryon antibaryon mass threshold and the widths are a few ten to
less than 200~MeV/$c^2$. Stimulated by recent experimental
results, a number of theoretical speculations and investigations
are put forth~\cite{th_gppb,th_cbbpair}, some focus on the
interpretation of a particular final state~\cite{th_gppb}, for
instance $\jpsi \ra \gamma \ppb$, while others discuss the final
states containing baryon and antibaryon pair~\cite{th_cbbpair}.
Most of these works devote to the improvement of previous models
or exploration of the production and decay dynamics, but it is
still far from understanding the problem.

The discovery of increased number of baryon-antibaryon enhancements 
near thresholds can not help reminding us of the era prior to the
development of $SU(3)$ quark model, when the so-called elementary 
particles emerged one by one. In this Letter, instead of studying the 
dynamics of the production or decay of these states, we try to find a 
way to classify them. We suggest a nonet scheme to accommodate the 
baryon-antibaryon enhancements observed in charmonium and $B$ decays. 
We surmise, with certain kind of interaction, for example, the residual 
strong force between the quarks inside the baryons, some multiplets can be formed as the baryon-antibaryon bound states. Our idea is enlightened by 
the Fermi-Yang-Sakata (FYS) model, in which the mesons were interpreted
as baryon-antibaryon bound states. In the following parts of the Letter, 
we first review the FYS model briefly, then put forward a
baryon-antibaryon nonet scheme, by virtue of which, many new
baryon-antibaryon bound states are expected. 
Finally, the search for these states are discussed.

\section{Fermi-Yang-Sakata Model}

In 1950s, as the number of the so-called elementary particles increased,
it became less likely that all of them were truly elementary. Under such
circumstance, as a tentative scheme, Fermi and Yang
proposed~\cite{fermi_yang} that the $\pi$-meson may be a composite
particle formed by the association of a nucleon and an
anti-nucleon, with strong attractive force in between which
binds them together. Since the mass of the $\pi$-meson is substantially
smaller than twice the mass of a nucleon, it is necessary to
assume that the binding energy is extremely large which is 
unappealing theoretically.

In 1955 after the discovery of the strangeness, Sakata extended
Fermi-Yang's idea by including a strange baryon $\Lambda$ and its
anti-particle~\cite{sakata}, and intended providing a physical
meaning for the Nishijima-Gell-Mann's rule~\cite{nishijima}. Four
years later, the most modern-like version of the FYS model was
developed by Ikeda, Ogawa and Ohnuki~\cite{ikeda}. They assumed
that proton $p$, neutron $n$ and $\Lambda$ are basic particles
which compose other baryons and mesons as suggested by the FYS
model. They proposed a framework which explicitly assures the
equivalence of the three basic particles, $p$, $n$ and $\Lambda$,
in the limit of an mass. This leads to the introduction of a new 
invariance under the exchange of $\Lambda$ and $p$ or $\Lambda$ and $n$ 
in addition to the usual charge independence and the conservation of 
electrical and hyperonic charge. They utilized $U(3)$ group to analyze the symmetry of the FYS model and obtained exactly the same classification of 
the pseudoscalar mesons as the quark model as long as the basic elements 
$p$, $n$ and $\Lambda$ are replaced by $u$, $d$ and $s$ quarks. 
The symmetry analysis of Ikeda, Ogawa and Ohnuki was so successful that 
all the pseudoscalar mesons known by 1961 could be accounted for, and
moreover, a new particle $\eta$ was predicted which was shortly
discovered~\cite{exp_eta}.

However, after the theory of unitary symmetry of strong interactions 
was put forward~\cite{gmnm}, especially when hyperon $\Omega^-$ was 
predicted definitely by Gell-Mann~\cite{gellman} and its existence was 
confirmed experimentally~\cite{exp_omega}, the FYS model became a history 
for the quark model. In fact, even when the FYS model was proposed, 
it encountered a profound difficulty which was the enormous binding energy for sticking the nucleons together to form a meson. On the contrary, 
for the newly observed baryon antibaryon enhancements near thresholds, 
the binding energy is small compared with the mass of a nucleon.
So we turn to the FYS model to classify these bound states.

%\section{Pseudoscalar and Vector Nonets}
\section{$0^-$ and $1^-$ Nonets}

We come back to the FYS model, but from a different point of view.
In our scheme, the baryon-antibaryon bound states do not refer to
ordinary mesons, such as $\pi$, $K$, $\eta$, but to 
the bound states formed by baryon and antibaryon. The interaction 
between the baryon and antibaryon is probably the residual force 
between the strong interaction of the quarks and gluons inside the baryon 
or antibaryon. On one hand, the masses of the three-quark systems 
(the baryon and the antibaryon) increase by a small amount
due to the residual forces 
required to form the bound state; on the other hand, the binding energy
between the two three-quark systems decreases the mass of the
baryon-antibaryon system to lower than the sum of the masses of
the three-quark systems, but very close to the baryon-antibaryon
mass threshold. This supplies as a phenomenological surmise, the real 
physics awaits for the validity of the quantum chromodynamics.

\begin{figure}[htb]
\centerline{\hbox{\psfig{file=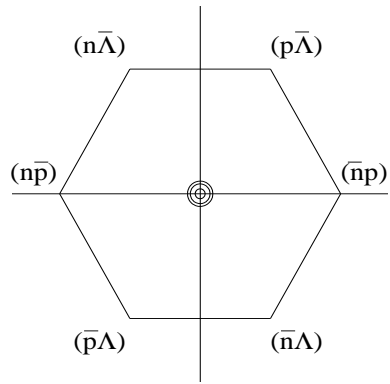,width=5.0cm,height=5.0cm}}}
\caption{\label{fig_nnt}Baryon-antibaryon nonet. The $J^P$ of
this nonet is either $0^{-}$ or $1^{-}$. The three circles in
the figure indicate the following three states: $(\nnb -
\ppb)/\sqrt{2}$, $(\nnb + \ppb -2 \LLb)/\sqrt{6}$, and $(\nnb +
\ppb + \LLb)/\sqrt{3}$. }
\end{figure}

Similar to the $SU(3)$ quark-antiquark nonets, we postulate the
existence of special octets and singlets (nonets) whose elements
are baryon-antibaryon bound states, as shown in Fig.~\ref{fig_nnt}. 
Hereafter, we limit our study to the low-mass baryons: $p$, $n$ and 
$\Lambda$. For a baryon-antibaryon bound state, its quantum numbers 
are obtained in the following way:
\begin{itemize}
\item Its spin ($S$) is 0 or 1, from the addition of the component baryons.
\item The parity is $(-1)^{L+1}$, where $L$
is the orbital angular momentum between the baryon and the antibaryon. 
In case of $S$-wave ($L=0$), the parity is odd ($-$), while for
$P$-wave, the parity is even ($+$). 
\item For pure neutral system,
such as $\nnb$, $\ppb$, or $\LLb$, the $C$-parity is %calculated by
$(-1)^{L+S}$. For charged members, we define the generalized 
$C$-parity~\cite{haber} by the neutral member of the nonet, but under 
$C$-parity transformation, the particle changes into its anti-particle.
\end{itemize}

\begin{table}[bth]
\caption{\label{tab_nnt}Quantum numbers of baryon-antibaryon nonet
with $J^P=0^{-}$ or $1^{-}$. $I$ is isospin, $I_3$ is the third
component of $I$, ${\cal S}$ is the strangeness and $Q$ is the 
charge of the state. The column ``symbol" gives nomenclature of the
$0^{-}$ and $1^{-}$ states for easy reference in the Letter.} {
%pseudoscalar and vector states for easy reference in the Letter.} {
\small\footnotesize
\begin{tabular}{clrrc} \hline \hline
 state & I($I_3$) &${\cal S}$&  Q  & Symbol               \\ \hline
$\bnbp$& 1($+1 $) & 0      &$+1$ &$\pi^+_B$/$\rho^+_B$  \\
$\bnpb$& 1($-1 $) & 0      &$-1$ &$\pi^-_B$/$\rho^-_B$  \\
${\displaystyle \frac{(\nnb-\ppb)}{\sqrt{2}}}$  {\rule[-2mm]{0mm}{7mm}}
       & 1($ 0 $)  & $ 0 $  &$0$  &$\pi^0_B$/$\rho^0_B$  \\
$\bplb$&$\frac{1}{2}$($+\frac{1}{2}$) & $+1 $  &$+1$ &$K^+_B/K^{*+}_B$      \\
$\bnlb$&$\frac{1}{2}$($-\frac{1}{2}$) & $+1 $  &$ 0$ &$K^0_B/K^{*0}_B$      \\
$\bnbl$&$\frac{1}{2}$($+\frac{1}{2}$) & $-1 $  &$ 0$
&$\overline{K}^0_B/
                                     \overline{K}^{*0}_B$ \\
$\bpbl$&$\frac{1}{2}$($-\frac{1}{2}$) & $-1 $  &$-1$ &$K^-_B/\overline{K}^{*-}_B$      \\
${\displaystyle \frac{(\nnb+\ppb-2\LLb)}{\sqrt{6}}}$ {\rule[-2mm]{0mm}{7mm}}
       & 0($ 0 $)  & $ 0 $  &$0$  &$\eta^8_B/\omega^8_B$ \\\hline
${\displaystyle \frac{(\nnb+\ppb+\LLb)}{\sqrt{3}}}$ {\rule[-2mm]{0mm}{7mm}}
       & 0($ 0 $)  & $ 0 $  &$0$  &$\eta^1_B/\omega^1_B$ \\  \hline \hline
\end{tabular} \\
}
\end{table}
%\end{table*}

The property of the $S$-wave spin-singlet states 
($J^{P}=0^{-}$) or the $S$-wave spin-triplet states 
($J^{P}=1^{-}$) nonet is summarized in Table~\ref{tab_nnt}, 
which leads to the relation of their production rates in experiments. 
For simplicity, we assign the particles in the nonets the same names as 
the meson nonets formed by quark-antiquark, but with a subscript $B$ for 
distinction. For example, the $0^-$ isospin vector states are denoted as 
$\pi^+_B$, $\pi^0_B$, and $\pi^-_B$, while the $1^-$ strange
particles are referred to as $K^{*0}_B$, $K^{*+}_B$, and their
anti-particles. In general, there is mixing between the 8th
component of the $SU(3)$ octet and the $SU(3)$ singlet. Since the
masses of proton and neutron differ by only 1.3~MeV$/c^2$, but the mass
of $\Lambda$ is 177~MeV$/c^2$ greater, one is invited to assume ideal
mixing between them, thus one is pure $(\ppb+\nnb)$ and the other is pure 
$\LLb$. This is to be verified by experiments.

If the electromagnetic interaction is neglected, the production
rates of these baryon-antibaryon bound states in $\jpsi$ and
$\psp$ decays can be simply related by $SU(3)$ symmetry except for a phase
space factor. For example, the production rates of
baryon-antibaryon bound states with $J^P=1^-$ accompanying by a 
pseudoscalar meson are related by: $\pi^0\rho^0_B : \pi^+\rho^-_B :
\pi^-\rho^+_B : K^+\overline{K}^{*-}_B : K^0\overline{K}^{*0}_B:
K^-K^{*+}_B : \overline{K^0}K^{*0}_B : \eta\omega_B : \eta\phi_B :
\eta^\prime\omega_B : \eta^\prime\phi_B = 1 : 1 : 1 : 1 : 1 : 1 :
1 : X^2_{\eta} : Y^2_{\eta} : X^2_{\eta^\prime} :
Y^2_{\eta^\prime}$. Here $X_\eta,~X_{\eta^\prime},~Y_\eta$, and
$Y_{\eta^\prime}$ are the mixing coefficients of $\eta$ and
$\eta^\prime$ \cite{etaxy}:
 \beqns
 |\eta\rangle        &=& X_\eta\frac{1}{\sqrt{2}} |u\overline{u}
 + d\overline{d} \rangle + Y_\eta |s\overline{s} \rangle~, \\
 |\eta^\prime\rangle &=& X_{\eta^\prime}\frac{1}{\sqrt{2}}
 |u\overline{u} + d\overline{d} \rangle + Y_{\eta^\prime}
 |s\overline{s} \rangle~,
 \eeqns
with $X_\eta=Y_{\eta^\prime}$, $X_{\eta^\prime}=-Y_\eta$, and
$X_\eta^2+Y_\eta^2=1$. Assuming aforementioned states predominantly 
decay to baryon-antibaryon pair, and considering the fact that
$\rho^0_B\to n \overline{n}$ and $\omega_B\to n \overline{n}$ are hard
to be detected experimentally, above relation can be reformulated
in terms of the experimentally detected final states:
$\pi^0(p\overline{p}) : \pi^+(n\overline{p}) :
\pi^-(\overline{n}p) : K^+(\overline{p}\Lambda) :
K^0(\overline{n}\Lambda): K^-(p\overline{\Lambda}) :
\overline{K^0}(n\overline{\Lambda}) : \eta(p\overline{p}) :
\eta(\Lambda\overline{\Lambda}) : \eta^\prime(p\overline{p}) :
\eta^\prime(\Lambda\overline{\Lambda}) \cong \frac{1}{2} : 1 : 1 :
1 : 1 : 1 : 1 : \frac{X^2_{\eta}}{2} : Y^2_{\eta} :
\frac{X^2_{\eta^\prime}}{2} : Y^2_{\eta^\prime}$. The production
rates of baryon-antibaryon bound states with other quantum numbers
are expressed similarly. The phase space is proportional to
$p^3$ for the production of the $J^P=1^-(0^-)$ baryon-antibaryon bound
state with an accompanying pseudoscalar (vector) meson, 
where $p$ is the momentum of the baryon-antibaryon bound state.

The $\ppb$ state observed in $\jpsi$ radiative
decays is $\eta_B$ or $\piz_B$ if it is a $S$-wave
state due to spin-parity conservation, and the
$p\overline{\Lambda}$ states in $\jpsi$ decays is $K^{*+}_B$ or
$K^+_B$. In $B$ decays, since parity is not conserved, 
the spin-parity of the state is to be determined by the angular 
distributions of the final state particles.

\section{Experimental Searches}

Because of the large phase space, $B$ decays play important roles in the 
study of the baryon-antibaryon resonances. Many of 
the baryon-antibaryon-pair-contained final states have been analyzed 
experimentally as mentioned above, other interesting modes to be searched 
for are given in Table~\ref{tab_bch}. The complexity here is the possible 
existence of two or more baryon-antibaryon resonances in the same final 
states and in a very small mass region, since many different $J^{P}$
states can be produced in $B$ decays, depending on the other
particles accompanying the baryon-antibaryon resonances.

%\begin{table*}[bth]
\begin{table}[htb]
\caption{\label{tab_bch}Possible decay modes
containing baryon-antibaryon nonets in $B$ decays.}
{\small %\footnotesize
\begin{tabular}{cl|cl} \hline \hline
           &~ decay mode     &         &~ decay mode   \\
                                                            \hline
$d\overline{b}\ra$
           & $d\overline{s}$  &$u\overline{b}\ra$
                                          & $u\overline{s}$  \\ \hline
$B^0 \ra$  & $\piz\bnlb$      &$B^+ \ra$  & $\pip\bnlb$       \\
           & $\pim\bplb$      &           & $\piz\bplb$       \\
           & $\kap\bnpb$      &           & $\kap\bppb$       \\
           & $\ks \bppb$      &           & $\kap\bllb$       \\
           & $\ks \bllb$      &           & $\ks \bnbp$       \\
           & $\eta\bnlb$      &           & $\eta\bplb$       \\
           & $\etap\bnlb$     &           & $\etap\bplb$      \\
           & $\roz\bnlb$      &           & $\rop\bnlb$       \\
           & $\rom\bplb$      &           & $\roz\bplb$       \\
           & $K^{*+}\bnpb$    &           & $K^{*+}\bppb$     \\
           & $K^{*0}\bppb$    &           & $K^{*+}\bllb$     \\
           & $K^{*0}\bllb$    &           & $K^{*0}\bnbp$     \\
           & $\omega\bnlb$    &           & $\omega\bplb$     \\
           & $\phi \bnlb$     &           & $\phi \bplb$      \\ \hline
$d\overline{b}\ra$
           &$d(\overline{c}c\overline{s})$
                              &$u\overline{b}\ra$
                                 & $u(\overline{c}c\overline{s})$ \\ \hline
$B^0 \ra$  & $\etac\bnlb$     &$B^+ \ra$  & $\etac\bplb$      \\
           & $\jpsi\bnlb$     &           & $\jpsi\bplb$      \\ \hline
$d\overline{b}\ra$
           &$d(\overline{c}u\overline{d})$
                              &$u\overline{b}\ra$
                                 & $u(\overline{c}u\overline{d})$ \\ \hline
$B^0 \ra$ & $\dzb\bppb$       &$B^+ \ra$  & $\dzb\bnbp$       \\
          & $\dzb\bllb$       &           &                   \\
          & $\dem\bnbp$       &           &                   \\
          &$\dszb (2007)\bppb$&           &$\dszb (2007)\bnbp$\\
          &$\dszb (2007)\bllb$&           &                   \\
          &$D^{*-}(2010)\bnbp$&           &                   \\
          &$D_s^- \bplb$      &           &                   \\
\hline \hline
\end{tabular} %%\\
}
\end{table}
%\end{table*}

Charmonium is another domain to study the baryon-antibaryon states. 
Unlike $B$ decays, conservation law holds a rein on a possible decay mode 
herein. By virtue of the quantum numbers listed in Table~\ref{tab_nnt}, 
some decay modes involving the $0^-$ and $1^-$ baryon-antibaryon bound 
states are listed in Table~\ref{tab_cch}.

%\begin{table*}[bth]
\begin{table}[bthp]
\caption{\label{tab_cch}Decay modes containing
baryon-antibaryon nonets in charmonium decays. The first $J^P$ is
for the accompanying particle while the second for the
baryon-antibaryon resonance.}
{\footnotesize%%\small
\begin{tabular}{clc} \hline \hline
             &~~~~~~~~decay mode                     & Note \\
                                                            \hline

$1^-\&0^{-}$  & $\roz\bppb,~\rop\bnpb,~\rom\bnbp$   &      \\
              & $K^{*+}\bpbl,~K^{*-}\bplb$          &$\ast$\\
              & $K^{*0}\bnbl,~\overline{K}^{*0}\bnlb$&$\ast$\\
              & $\omega\bppb$                       &      \\
              & $\phi\bllb$                         &$\ast$\\
 $1^+\&0^{-}$ & $b_1^0(1235)\bppb$                  &$\ast$\\
              & $b_1^+(1235)\bnpb,~b_1^-(1235)\bnbp$&$\ast$\\
              & $h_1(1170)\bppb,~h_1(1170)\bllb,~$  &$\ast$\\
              & $K_1^{+}(1270)\bpbl,~K_1^{-}(1270)\bplb$
                                                    &$\ast$\\
              & $K_1^{0}(1270)\bnbl,~
                 \overline{K}_1^{0}(1270)\bnlb$     &$\ast$\\
              & $K_1^{+}(1400)\bpbl,~K_1^{-}(1400)\bplb$
                                                    &$\ast$\\
              & $K_1^{0}(1400)\bnbl,~
                 \overline{K}_1^{0}(1400)\bnlb$     &$\ast$\\
 \hline
 $0^-\&1^{-}$ & $\piz\bppb,~\pip\bnpb,~\pim\bnbp$   &      \\
              & $\kap\bpbl,~\kaz\bnbl$              &      \\
              & $\kam\bplb,~\kzb\bnlb$              &      \\
              & $\eta\bppb,~\eta\bllb$              &      \\
              & $\etap\bppb,~\etap\bllb$            &$\ast$\\
 $0^+\&1^{-}$ & $a_0^0(980)\bppb$                   &      \\
              & $a_0^+(980)\bnpb,~a_0^-(980)\bnbp$  &      \\
              & $a_0^0(1450)\bppb$                  &$\ast$\\
              & $a_0^+(1450)\bnpb,~a_0^-(1450)\bnbp$&$\ast$\\
              & $f_0(980)\bppb,~f_0(980)\bllb,$     &      \\
              & $f_0(1370)\bppb,~f_0(1370)\bllb,$   &$\ast$\\
              & $\ko^{*+}(1430)\bpbl,~\ko^{*-}(1430)\bplb$
                                                    &$\ast$\\
              & $\ko^{*0}(1430)\bnbl,~
                 \overline{\ko}^{*0}(1430)\bnlb$    &$\ast$\\
 $1^+\&1^{-}$ & $a_1^0(1260)\bppb$                  &$\ast$\\
              & $a_1^+(1260)\bnpb,~a_1^-(1260)\bnbp$&$\ast$\\
              & $f_1(1285)\bppb,~f_1(1420)\bllb,$   &$\ast$\\
              & $K_1^{+}(1270)\bpbl,~K_1^{-}(1270)\bplb$
                                                    &$\ast$\\
              & $K_1^{0}(1270)\bnbl,~
                 \overline{K}_1^{0}(1270)\bnlb$     &$\ast$\\
              & $K_1^{+}(1400)\bpbl,~K_1^{-}(1400)\bplb$
                                                    &$\ast$\\
              & $K_1^{0}(1400)\bnbl,~
                 \overline{K}_1^{0}(1400)\bnlb$     &$\ast$\\
 $2^+\&1^{-}$ & $a_2^0(1320)\bppb$                  &$\ast$\\
              & $a_2^+(1320)\bnpb,~a_2^-(1320)\bnbp$&$\ast$\\
              & $f_2(1270)\bppb,$                   &$\ast$\\
              & $f^{\prime}_2(1525)\bllb$           &$\ast \ast$\\
              & $K_2^{*+}(1430)\bpbl,~K_2^{*-}(1430)\bplb$
                                                    &$\ast$\\
              & $K_2^{*0}(1430)\bnbl,~
                 \overline{K}_2^{*0}(1430)\bnlb$    &$\ast$\\
\hline \hline
\end{tabular} \\
$\ast$: not allowed in $\jpsi$ decays;\\
$\ast$ $\ast$: not allowed in $\psp$ decay.}
\end{table}
%\end{table*}

The production of the $0^-$ baryon-antibaryon bound states in 
$\jpsi$ (or $\psp$) decays can be accompanied by a vector meson. For 
the iso-vector bound states, one may look for the $\rho N \overline{N}$ 
(nucleon antinucleon) final states,
including $\rho^+ n \overline{p}$, $\rho^0 p \overline{p}$ and
$\rho^- p \overline{n}$; for the iso-scalar bound state, one may look for
the $\omega p \overline{p}$ final state; while for the strange
states, one may look for the $K^{*+} \Lambda \overline{p} +
c.c.$ and $K^{*0} \Lambda \overline{n} + c.c.$ final states. The neutron or
anti-neutron which is not detected may be %measured by
reconstructed by kinematic fit in the event selection. %reconstruction. 
The $SU(3)$ singlet state can be searched for by measuring $\phi \LLb$ 
final state. The measurement of the $0^-$ baryon-antibaryon bound states 
together with an axial-vector meson is less promising since almost all the
axial-vector mesons are resonances.

The production of the $1^-$ baryon-antibaryon bound states can be
accompanied by a pseudoscalar ($\pi$, $\eta$, $\eta'$, $K$),
scalar, tensor or axial-vector meson. The most promising way 
to look for them is in the decays with a pseudoscalar meson:
analyze $\pi N \overline{N}$ for the iso-vector bound states; 
analyze $\eta p \overline{p}$ for iso-scalar bound state; 
and analyze $K^{+} \Lambda \overline{p} + c.c.$ and $K^{0} \Lambda
\overline{n} + c.c.$ for the strange bound states. The $SU(3)$ singlet
bound state can be searched for via $\eta' \LLb$.

It should be noted that the neutral non-strange $0^-$ baryon-antibaryon 
bound states can also be produced via radiative decays of $\jpsi$ (or
$\psp$), while the $1^-$ baryon-antibaryon bound states can not be 
produced this way due to spin-parity conservation.

Although among charmonium decays $\jpsi$ provides a good source of the
baryon-antibaryon bound states because of the large data samples,
there are disadvantages: the phase space is too small and there are 
many $N^*$'s near nucleon meson mass threshold, which affect the 
identification of the states. The $\psp$ decays have larger phase space, 
however, the data samples are smaller, and there is a large fraction of 
charmonium transition. CLEOc and BESIII will surely help to improve the 
%statistics, and a partial wave analysis (PWA) is desirable to 
statistics, and the partial wave analysis is desirable to 
take the $N^*$ contribution into account correctly.

It is also possible to perform such searches in
bottomonium ($\Upsilon$) decays, with the existing data sample at
CLEO-III and possibly more if $B$-factories take data at
$\Upsilon(1S)$. The phase space is much larger than in 
charmonium, and the $N^*$ states are far from the
baryon-antibaryon mass threshold. In principle, all modes listed
in Table~\ref{tab_cch} can be searched for in bottomonium decays.

\mbox{} \\
\section{Discussion and Conclusion}

Although our discussion is limited to $S$-wave, $SU(3)$
baryon-antibaryon bound states, it can be easily extended 
in many aspects. First, the $P$-wave, $D$-wave and even higher angular
momentum multiplets are also expected to exist. Thus we
have $J^P=0^+$, $1^+$, $2^+,\cdots$, states. Second, the scheme can be
extended by including more baryons, for example, the charmed
baryon $\Lambda_c$. As has been reported by the Belle
collaboration, an enhancement was observed in $\Lambda_c \bar{p}$
mass spectrum near the threshold~\cite{bellelcp}. This can be
interpreted as a member in the $SU(4)$ multiplets. 
Last, the extension to the baryon-meson, or the meson-meson bound states is
in principle straightforward. Nevertheless, the existence of such
kinds of resonances can merely be determined by experiment.

In summary, the observations of the enhancements near the
baryon antibaryon mass thresholds in charmonium and $B$ decays bring us
fresh ideas in the study of hadron spectroscopy. With the known
symmetry properties of strong interaction, we foresee the
existence of the nonet baryon-antibaryon bound states and their
possible quantum numbers in a revived FYS model, even though we cannot 
calculate the binding energies from the first principle of QCD at present. 
In the light of our scheme, we know where to find these bound 
states in $\jpsi$, $\psp$, $\Upsilon$ and $B$ meson decays. The
search can be conducted with the existing or soon available
CLEOc, BESIII, and $B$-factory data.

\end{document}